\documentclass[11pt,a4paper]{article}

\usepackage[margin=1in]{geometry}
\usepackage{amsmath,amssymb,amsfonts,amsthm,mathtools,bm}
\usepackage{braket}
\usepackage{booktabs}
\usepackage{microtype}
\usepackage{hyperref}
\usepackage{enumitem}
\usepackage{graphicx}
\usepackage{caption}
\hypersetup{colorlinks=true,linkcolor=blue,citecolor=blue,urlcolor=blue}
\newcommand{\OO}{\mathcal O}
\newcommand{\dd}{\mathrm d}
\newcommand{\Tr}{\operatorname{Tr}}
\newcommand{\sgn}{\operatorname{sgn}}
\newcommand{\sech}{\operatorname{sech}}
\newcommand{\Pc}{\mathcal P}

\newtheorem{proposition}{Proposition}

\title{\textbf{The Junction That Remembers:\\
Infrared Response of Two Coupled SYK Majorana Systems}}
\author{Ali Vahedi\thanks{Email: \texttt{vahedi@khu.ac.ir}}\\
Department of Physics, Kharazmi University, Tehran, Iran}

\date{}

\begin{document}
\maketitle

\begin{abstract}
Two strongly interacting Majorana systems of the Sachdev--Ye--Kitaev type, each with its own independent disorder, are joined by a link that is switched on in time. We ask how the infrared structure of the link is imprinted on the response and whether the system remembers how it was opened. At large $N$ the leading cross-correlation between the two systems is an exact product of two equilibrium propagators and the switching profile; because the disorder is independent, no ladder dresses it. The link operator has scaling dimension one half, which gives its susceptibility a universal logarithm whose coefficient and phase we fix and verify numerically, while the gap of the coupled equilibrium problem grows more slowly than the naive crossover estimate. Exact dynamics for up to twelve Majoranas per side show strong parity transfer and entanglement several times the free-fermion ceiling. The work done by the switch has a first-order part that does not depend on the switching time and a second-order part that grows in magnitude with it. Once the work is held fixed, the protocol dependence of the entanglement, the link expectation value and a scrambling diagnostic nearly vanishes, and the small residuals do not depend on whether the protocols differ at low or at high frequency. These small systems have no infrared window, so they neither confirm nor exclude memory at large $N$.
\end{abstract}

\tableofcontents
\bigskip

%=====================================================================
\section{Introduction}
\label{sec:intro}

A time-dependent coupling between two quantum systems changes the connectivity of their operator algebras. It opens a channel through which correlations, quantum information and operator support can flow from one system into the other. When each system is itself strongly interacting, the response of one is dressed by the dynamics of the other, and the question of what the junction remembers about the way it was opened becomes nontrivial.

The Sachdev--Ye--Kitaev (SYK) model provides a controlled setting for this question~\cite{SachdevYe,KitaevSYK,MaldacenaStanford,PolchinskiRosenhaus,Chowdhury,Saha2026}. It is a system of $N$ Majorana fermions with random all-to-all quartic interactions. It is solvable at large $N$, has an emergent conformal symmetry at low energy, is maximally chaotic~\cite{MaldacenaShenkerStanford}, and has a gravitational description in terms of nearly-$\mathrm{AdS}_2$ dynamics~\cite{MSY,KitaevSuh}. At finite $N$ it is also small enough to be diagonalized exactly.

Two SYK systems coupled by a bilinear link are the basis of the eternal traversable wormhole~\cite{MaldacenaQi}, and thermal and driven versions of this construction have been studied~\cite{LenskyQi,BerenguerFloquet,BerenguerHot}. In the Maldacena--Qi construction the two copies share the same disorder realization, so the low-energy dynamics are governed by a shared Schwarzian mode. Quenches and drives of a single SYK system are also well studied~\cite{Eberlein,Polkovnikov,Bhattacharya}. Less attention has been given to the case where the two copies are statistically \emph{independent} and the link is switched on in time. There the disorder average generates no interaction vertex between the systems, and the leading response is not dressed by the four-time ladder, a structural fact that sharply distinguishes this problem from the wormhole construction.

This is the setting of the present paper. We consider two $q=4$ SYK systems with independent disorder and couple them through a Majorana bilinear whose strength $g(t)$ is turned on smoothly from a small value $g_{\rm cl}$ to a larger value $g_{\rm op}$. We call the protocol a \emph{shutter}. There is no spatial aperture in SYK; what opens is the interaction graph, which changes from two nearly independent operator algebras to a single coupled one. The question we pursue is whether the infrared structure of the link leaves a lasting imprint, a \emph{memory}, of the way the shutter was opened, beyond what is accounted for by the energy it deposits.

Three levels of description are involved, and part of the purpose of this paper is to keep them apart. At finite $N$ the system is an ordinary closed quantum system with a time-dependent Hamiltonian and an exactly unitary evolution. At large $N$ the disorder-averaged dynamics obey coupled Schwinger--Dyson equations on the Keldysh contour. And one may ask whether some projection of the response can be summarized by an effective Bogoliubov-type kernel, as it can for a free Majorana shutter~\cite{Vahedi2026}. The last question is much more restrictive than the first two and must not be confused with the exact interacting dynamics.

For scrambling and operator growth we use the standard tools: the squared commutator and out-of-time-order correlator~\cite{MaldacenaShenkerStanford}, and the operator-size picture of SYK~\cite{RobertsStanfordStreicher}. We use the out-of-time-order correlator only as a qualitative diagnostic of cross-system operator sensitivity and extract no Lyapunov exponent, because the accessible sizes are too small for an exponential regime.

\paragraph{Summary of results.} At large $N$ and to first order in the link, the cross-correlation between the two systems is the product of two equilibrium propagators and the switching function (Sec.~\ref{sec:response}), with an overall factor of $i$ that we fix by comparison with exact diagonalization. Because the disorder ensembles are independent, no ladder dresses this response; the ladder acts on the diagonal correlators at second order (Sec.~\ref{sec:ladder}). The link operator has scaling dimension $1/2$, so it is relevant and its susceptibility has a logarithmic infrared term whose coefficient, sign and phase we determine and verify against the Euclidean large-$N$ equations. The coupled equilibrium problem has a gap that, in the accessible window, grows more slowly than the dimensional estimate $g^2/J$ (Sec.~\ref{sec:scales}). In exact dynamics for up to twelve Majoranas per side (Sec.~\ref{sec:finiteN}), parity transfer is strong and the entanglement entropy reaches several times the free-fermion ceiling. The work has a switching-time-independent first-order part and a second-order part that grows in magnitude with the switching time (Sec.~\ref{sec:worktau}). At equal work, the dependence on the protocol nearly disappears for the entropy, the link expectation value and the scrambling diagnostic (Secs.~\ref{sec:memory}--\ref{sec:otocmemory}), and the small residuals are the same whether the protocols differ at low or at high frequency (Sec.~\ref{sec:tuned}). We read this as a limitation of small systems, which have no infrared window, and not as evidence against memory at large $N$.

\paragraph{Organization.} Section~\ref{sec:model} defines the model, the protocol and every observable. Section~\ref{sec:largeN} introduces the large-$N$ contour equations. Section~\ref{sec:response} derives the first-order response, the junction susceptibility and the structure of the work. Section~\ref{sec:ladder} explains where the interacting ladder enters, and Sec.~\ref{sec:benchmark} analyzes a conditional kernel. Section~\ref{sec:scales} treats infrared scales. Section~\ref{sec:finiteN} presents the exact finite-size dynamics. Section~\ref{sec:discussion} discusses the physical picture and its limits, and Sec.~\ref{sec:conclusions} concludes. Technical details are in the appendices.

%=====================================================================
\section{Model, protocol and observables}
\label{sec:model}

We begin by stating the model and the quantities that will be measured. Everything in later sections is built from the definitions below, and Table~\ref{tab:symbols} collects the notation.

\subsection{Two SYK systems and a bilinear link}
\label{sec:hamiltonian}

Each system $a=L,R$ consists of $N$ Majorana fermions $\chi^a_i$ ($i=1,\dots,N$) obeying
\begin{equation}
\{\chi_i^a,\chi_j^b\}=\delta_{ij}\delta^{ab},
\label{eq:CAR}
\end{equation}
so that $(\chi_i^a)^2=1/2$. Each system has the $q=4$ SYK Hamiltonian
\begin{equation}
H_a=\sum_{1\le i<j<k<l\le N} J^a_{ijkl}\,\chi_i^a\chi_j^a\chi_k^a\chi_l^a,
\qquad
\overline{J^a_{ijkl}}=0,\quad
\overline{\left(J^a_{ijkl}\right)^2}=\frac{3!\,J^2}{N^3},
\label{eq:SYK}
\end{equation}
where the overline denotes disorder averaging and $J$ sets the energy scale. The couplings of the two systems are drawn \emph{independently}. The systems are joined by the link operator
\begin{equation}
\OO=i\sum_{i=1}^{N}\chi_i^L\chi_i^R,
\qquad
H(t)=H_L+H_R+g(t)\,\OO .
\label{eq:Htotal}
\end{equation}
The operator $\OO$ is Hermitian and extensive. Because it pairs flavor $i$ of one system with flavor $i$ of the other, it can be viewed as a chemical potential for the complex fermions $c_i=(\chi_i^L+i\chi_i^R)/\sqrt2$, since $\OO=\sum_i(c_i^\dagger c_i-\tfrac12)$. This observation is used in Sec.~\ref{sec:scales}.

The closed system is finite-dimensional and unitary. The word ``open'' therefore does not refer to a bath: it refers to the opening of a communication channel between two initially weakly connected many-body systems. Throughout, all analytic statements refer to the normalization in Eq.~\eqref{eq:Htotal}. Conventions in which the link carries an explicit factor $N^{-1/2}$ change the large-$N$ scaling of the deformation and are not equivalent.

\subsection{The switching protocol}
\label{sec:protocol}

The coupling is switched on smoothly,
\begin{equation}
g(t)=g_{\rm cl}+\frac{\delta g}{2}\left[1+\tanh\frac{t-t_0}{\tau}\right],
\qquad \delta g=g_{\rm op}-g_{\rm cl},
\label{eq:protocol}
\end{equation}
so that $g\to g_{\rm cl}$ at early and $g\to g_{\rm op}$ at late times. The opening happens around $t_0$ and takes a time of order $\tau$. A short $\tau$ (a ``sudden'' opening) injects high frequencies; a long $\tau$ (an adiabatic opening) does not. In a gapless conformal theory, however, a slow opening does not remove the response of arbitrarily soft modes, which is the origin of the infrared effects discussed below.

\begin{figure}[t]
\centering
\includegraphics[width=0.92\linewidth]{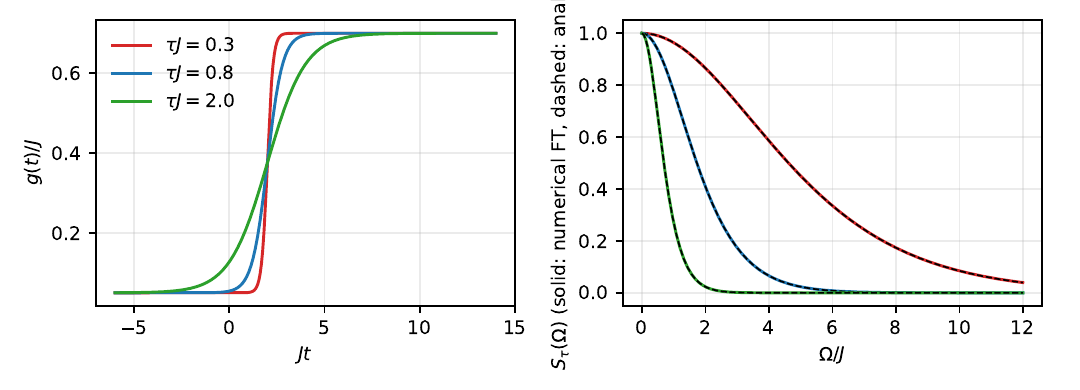}
\caption{The switching protocol. Left: $g(t)$ from Eq.~\eqref{eq:protocol} with $g_{\rm cl}/J=0.05$, $g_{\rm op}/J=0.70$ and $Jt_0=2$, for $\tau J=0.3,0.8,2.0$. Right: the filter $S_\tau(\Omega)$; the numerical Fourier transform (solid) coincides with Eq.~\eqref{eq:Stau} (dashed).}
\label{fig:protocol}
\end{figure}

The frequency content of the switching is controlled by $\dot g$, which is localized in time:
\begin{equation}
\dot g(t)=\frac{\delta g}{2\tau}\,\sech^2\frac{t-t_0}{\tau}.
\label{eq:gdot}
\end{equation}
With the Fourier convention $\widetilde f(\Omega)=\int\dd t\,e^{i\Omega t}f(t)$, its transform is
\begin{equation}
\widetilde{\dot g}(\Omega)=\delta g\,e^{i\Omega t_0}\,S_\tau(\Omega),
\qquad
S_\tau(\Omega)=\frac{\pi\tau\Omega/2}{\sinh(\pi\tau\Omega/2)} .
\label{eq:Stau}
\end{equation}
The filter $S_\tau$ equals one at $\Omega=0$ and decays exponentially for $|\Omega|\tau\gg1$: a protocol of duration $\tau$ excites only frequencies below about $1/\tau$. We checked Eq.~\eqref{eq:Stau} by numerical Fourier transform for the three values $\tau J=0.3,0.8,2.0$ used below; the maximum deviation is $4.5\times10^{-16}$ (Fig.~\ref{fig:protocol}). Working with $\dot g$ rather than $g$ avoids boundary terms, because $g$ itself does not vanish at late times.

\subsection{Work}
\label{sec:work}

Because only $H_{\rm link}$ depends on time, the energy of any state evolving under Eq.~\eqref{eq:Htotal} changes at the rate
\begin{equation}
\frac{\dd}{\dd t}\langle H(t)\rangle=\dot g(t)\,\langle\OO(t)\rangle,
\qquad
W=\int_{t_i}^{t_f}\dd t\;\dot g(t)\,\langle\OO(t)\rangle
=\langle H(t_f)\rangle-\langle H(t_i)\rangle .
\label{eq:work}
\end{equation}
The total work $W$ done by the protocol is therefore a measurable energy difference. This identity is exact at any $N$. It separates energy injection from the transfer of correlations: entanglement and scrambling may change without being determined solely by $W$.

\subsection{Observables}
\label{sec:observables}

The finite-size study of Sec.~\ref{sec:finiteN} follows four kinds of observable, defined here once.

\emph{Left parity.} The fermion parity of the left system is $P_L=\prod_{k=1}^{N/2}(-2i\,\chi^L_{2k-1}\chi^L_{2k})$, with eigenvalues $\pm1$. We follow the normalized parity $p_L(t)=\langle P_L(t)\rangle/\langle P_L(t_i)\rangle$, which starts at $1$. It measures how much of an initial left-parity excitation has moved to the right system. Total parity $P_LP_R$ is conserved.

\emph{Entanglement.} With $\rho_L(t)=\Tr_R|\Psi(t)\rangle\langle\Psi(t)|$, the left entropy is $S_L=-\Tr\rho_L\ln\rho_L$. It is bounded by $\ln\dim\mathcal H_L=\tfrac N2\ln2$.

\emph{Cross-system scrambling.} We use the infinite-temperature out-of-time-order correlator
\begin{equation}
F_{LR}(t)=\frac{1}{\dim\mathcal H}\Tr\!\left[V(t)\,W\,V(t)\,W\right],
\qquad V=\gamma^L_1,\; W=\gamma^R_1,
\label{eq:OTOC}
\end{equation}
where $\gamma=\sqrt2\chi$ has $\gamma^2=1$ and $V(t)$ is the Heisenberg operator evolved with the full time-dependent unitary from $t_i$. For anticommuting Majoranas $F_{LR}(t_i)=-1$, and we plot $-F_{LR}$, which starts at $1$. It is a property of the Hamiltonian, not of the initial state.

\emph{Memory functional.} Given two protocols with the same initial and final couplings, labeled $1$ and $2$, the memory functional of an observable $X$ at a late time $t_m$ is
\begin{equation}
\mathcal M_X(1,2;t_m)=X(t_m;1)-X(t_m;2).
\label{eq:memory}
\end{equation}
It vanishes if the late-time value of $X$ does not depend on how the coupling was opened. Since the protocols generally inject different amounts of work, a meaningful test compares protocols at \emph{equal} $W$ (Sec.~\ref{sec:memory}).

\begin{table}[t]
\centering
\small
\begin{tabular}{ll}
\toprule
Symbol & Meaning\\
\midrule
$N$, $J$ & Majoranas per system; SYK energy scale\\
$\chi^a_i$, $\gamma^a_i$ & Majorana operators, $\chi=\gamma/\sqrt2$; $a=L,R$\\
$\OO$, $g(t)$ & link operator; link strength\\
$g_{\rm cl}$, $g_{\rm op}$, $t_0$, $\tau$ & closed and open couplings, opening time and duration\\
$S_\tau(\Omega)$ & switching filter, Eq.~\eqref{eq:Stau}\\
$G$, $G^R$, $G_c$ & Green function, retarded part, conformal limit\\
$\circ$ & convolution over the Keldysh contour\\
$\Delta_\chi$, $\Delta_\OO$ & scaling dimensions ($1/4$, $1/2$)\\
$\chi^R_{\OO\OO}$ & retarded susceptibility of $\OO$\\
$E_g$ & dimensional crossover scale, $\sim g^2/J$\\
$p_L$, $S_L$, $F_{LR}$ & parity, entropy, OTOC (Sec.~\ref{sec:observables})\\
$W$, $\mathcal M_X$ & work, memory functional\\
$\mu$, $\omega_\ast$ & infrared cutoff and conformal upper scale\\
\bottomrule
\end{tabular}
\caption{Notation.}
\label{tab:symbols}
\end{table}

%=====================================================================
\section{Large-\texorpdfstring{$N$}{N} description}
\label{sec:largeN}

At large $N$ the disorder average reduces the problem to a pair of self-consistent equations for the two-point function and the self-energy. We set these up on the closed-time (Keldysh) contour, which is the right framework when $g$ depends on time.

\subsection{Contour equations}
\label{sec:contour}

Define the flavor-averaged contour Green function
\begin{equation}
G_{ab}(t,t')=-\frac{i}{N}\sum_{i=1}^N\left\langle T_C\,\chi^a_i(t)\,\chi^b_i(t')\right\rangle,
\qquad a,b=L,R,
\label{eq:Gcontour}
\end{equation}
where $T_C$ orders operators along the contour. With this normalization the free retarded function is $G^R=-i\theta(t-t')$, and in general $G^R_{ab}(t,t')=-i\theta(t-t')\,\overline{\langle\{\chi^a_i(t),\chi^b_i(t')\}\rangle}$. At large $N$ the Green functions obey the Dyson equation
\begin{equation}
\left[i\partial_t\,\delta_{ac}-\Sigma_{ac}-\Sigma^{\rm link}_{ac}\right]\circ G_{cb}=\delta_{ab}\,\delta_C,
\label{eq:Dyson}
\end{equation}
where $\circ$ is convolution along the contour and $\delta_C$ the contour delta function. Averaging over the disorder in Eq.~\eqref{eq:SYK} yields
\begin{equation}
\Sigma_{LL}=J^2\,G_{LL}^{\,3},\qquad \Sigma_{RR}=J^2\,G_{RR}^{\,3},\qquad \Sigma^{\rm SYK}_{LR}=\Sigma^{\rm SYK}_{RL}=0 .
\label{eq:selfdiag}
\end{equation}
The vanishing of the off-diagonal SYK self-energy is the key structural fact of this paper. It holds because the couplings of the two systems are independent, so the average $\overline{J^LJ^R}$ vanishes and no interaction line connects the two systems. The only object that couples them is the link, whose self-energy is local in time,
\begin{equation}
\Sigma^{\rm link}_{LR}(t,t')=i\,g(t)\,\delta_C(t,t'),
\qquad
\Sigma^{\rm link}_{RL}(t,t')=-i\,g(t)\,\delta_C(t,t').
\label{eq:linkself}
\end{equation}
The factor $i$ follows from the Heisenberg equations of motion $\dot\chi^L_i=i[H_L,\chi^L_i]+g\,\chi^R_i$ and $\dot\chi^R_i=i[H_R,\chi^R_i]-g\,\chi^L_i$, and it makes the link self-energy Hermitian. We confirm this factor against exact diagonalization in Sec.~\ref{sec:prop1num}. After the usual rotation to retarded, advanced and Keldysh components (Appendix~\ref{app:keldysh}), Eq.~\eqref{eq:Dyson} becomes a matrix equation in both flavor and Keldysh space.

\subsection{The decoupled conformal saddle}
\label{sec:saddle}

At $g=0$ the two systems are independent and $G_{ab}=\delta_{ab}G$, with $G$ the equilibrium propagator of a single SYK system. In the infrared window $|\omega|\ll J$ (but above any temperature or finite-size scale), and with the Euclidean convention $G(\tau)=\overline{\langle T_\tau\chi_i(\tau)\chi_i(0)\rangle}$ for which $G\to\tfrac12\sgn\tau$ at $J=0$, the solution is conformal,
\begin{equation}
G_c(\tau)=b\,\frac{\sgn\tau}{(J|\tau|)^{1/2}},
\qquad
b=(4\pi)^{-1/4},
\qquad
\Delta_\chi=\frac14 .
\label{eq:Gc}
\end{equation}
The amplitude $b$ follows from the standard $q=4$ condition $\pi b^4=1/4$ for the normalization of Eqs.~\eqref{eq:CAR} and~\eqref{eq:SYK}~\cite{MaldacenaStanford}, and the exponent from the scaling dimension $\Delta_\chi=1/q$. In frequency space the retarded and advanced functions scale as $G^{R/A}_c(\omega)\sim J^{-1/2}|\omega|^{-1/2}$. A Euclidean propagator, a retarded function, a spectral density and a Keldysh component are different objects; below, $G_c$ denotes the component appropriate to the equation in which it appears.

%=====================================================================
\section{First-order response and the junction susceptibility}
\label{sec:response}

We now ask how the two systems respond to the switching, to first order in $g$. Two complementary descriptions are available. One follows the cross-correlation between the systems, which is zero before the link is opened and is generated by it. The other follows the response of the link operator itself, which controls the energy absorbed.

\subsection{The cross-correlation to first order}
\label{sec:prop}

Expand $G=G^{(0)}+\delta G+O(g^2)$ about the decoupled saddle. At first order the only term that connects the two systems is the link self-energy, Eq.~\eqref{eq:linkself}, because $\Sigma^{\rm SYK}_{LR}=0$. Linearizing the Dyson equation and multiplying by two equilibrium propagators gives the following.

\begin{proposition}\label{prop:leading}
To first order in the time-dependent link, about the decoupled large-$N$ saddle,
\begin{equation}
\boxed{\;\delta G_{LR}=i\;G\circ g\circ G\;}
\label{eq:mainresponse}
\end{equation}
where $G$ is the equilibrium propagator of a single SYK system and $g$ acts as a multiplicative, time-local source.
\end{proposition}

\begin{proof}
At $g=0$ the inverse propagator is block diagonal, $G_0^{-1}=\mathrm{diag}(G^{-1},G^{-1})$. The perturbation of the inverse propagator is the link self-energy, which is off-diagonal and local in time. Since $\Sigma^{\rm SYK}_{LR}=0$, nothing else feeds the $LR$ block at this order, and $\delta G=G_0\,\Sigma^{\rm link}\,G_0$ gives Eq.~\eqref{eq:mainresponse}. The diagonal blocks change only at $O(g^2)$.
\end{proof}

Two features deserve emphasis. First, Eq.~\eqref{eq:mainresponse} contains the \emph{exact} equilibrium propagator of each system, not only its conformal limit; the conformal form $G_c$ is recovered inside the conformal window. Second, it is not a Bogoliubov transformation: it is a statement about a correlation function, and it holds for the interacting system. The prefactor can be checked in the free limit $J\to0$, where the equations of motion can be solved exactly. There $G^R_{LR}(t,t')=-i\theta(t-t')\sin\!\big[\int_{t'}^{t}g\big]$, whose first-order term is $-i\theta(t-t')\int_{t'}^{t}g$, exactly $i\,G^R\circ g\circ G^R$ with $G^R=-i\theta$.

\subsection{Exact test of the first-order response}
\label{sec:prop1num}

\begin{figure}[t]
\centering
\includegraphics[width=0.92\linewidth]{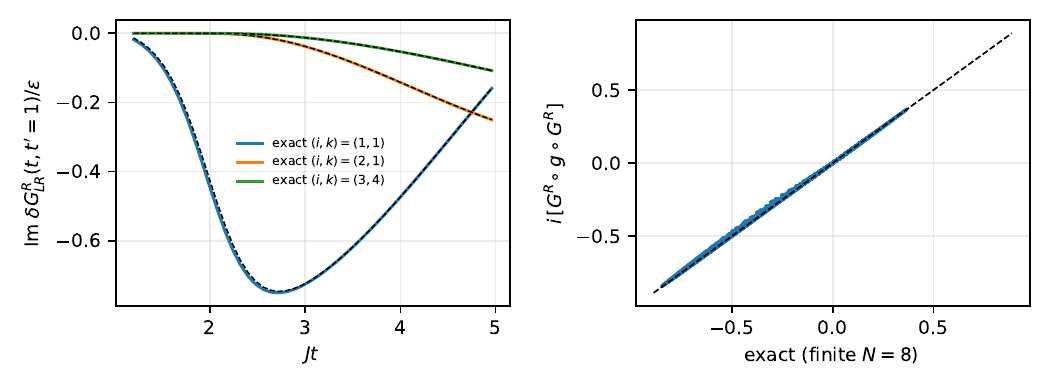}
\caption{Exact finite-size test of Proposition~\ref{prop:leading} ($8+8$ Majoranas, one disorder sample, all $8\times8$ flavor pairs). Left: imaginary part of $\delta G^R_{LR}(t,t'{=}1)/\varepsilon$ for three flavor pairs and the prediction $i\,[G^R\!\circ g\circ G^R]$ (dashed). Right: all elements, exact against prediction; the dashed line is the identity.}
\label{fig:prop1}
\end{figure}

Proposition~\ref{prop:leading} is an identity at large $N$, but it can be tested at finite $N$ because, for a product initial state of two independent systems, the first-order cross-correlation factorizes exactly into the two exact single-system propagators. For one disorder realization of the $8+8$ Majorana system we computed the exact retarded matrix $G^R_{LR,ik}(t,t')$ for a weak pulse $g(t)=\varepsilon\,\sech^2[(t-2)/0.5]$ and compared it with the convolution of the exact single-system matrices $G^R_{L,ij}$ and $G^R_{R,jk}$ over all $8\times8$ flavor pairs. With the prefactor $i$ the relative error is $2.1\%$, against $140\%$ for $\pm1$ and $199\%$ for $-i$ (Fig.~\ref{fig:prop1}). The response is linear in $\varepsilon$ to $1.7\times10^{-4}$ between $\varepsilon=0.01$ and $0.02$, and the residual is consistent with the trapezoid quadrature and $O(\varepsilon^2)$ terms. This fixes the sign and the factor $i$, and it confirms that no ladder correction enters at this order.

\subsection{Two-frequency form}
\label{sec:twofreq}

Because the source depends on time, the response is not stationary and depends on two frequencies. With
\begin{equation}
G(\omega,\nu)=\int\dd t\,\dd t'\,e^{i\omega t-i\nu t'}\,G(t,t'),
\label{eq:twofreqconv}
\end{equation}
Eq.~\eqref{eq:mainresponse} reads
\begin{equation}
\delta G_{LR}(\omega,\nu)=i\,G(\omega)\,\widetilde g(\omega+\nu)\,G(\nu).
\label{eq:twosource}
\end{equation}
The sum $\omega+\nu$ is the frequency supplied by the nonstationary source, and $\widetilde g(\omega+\nu)$ carries the switching history through the filter $S_\tau$ of Eq.~\eqref{eq:Stau}. The remaining factors are the infrared propagators of the two systems. This is the controlled origin of the $\omega+\nu$ structure that appears in effective-kernel constructions (Sec.~\ref{sec:benchmark}).

\subsection{The junction susceptibility}
\label{sec:susc}

The second description follows the link operator. At the decoupled saddle, Wick factorization gives
\begin{equation}
\frac1N\left\langle\OO(\tau)\OO(0)\right\rangle_0=G(\tau)^2 .
\label{eq:OOwick}
\end{equation}
No vertex correction contributes at leading order, because no interaction line connects the two systems; the product of two exact propagators is the full large-$N$ result. In the conformal window this gives a universal power law,
\begin{equation}
\frac1N\left\langle\OO(\tau)\OO(0)\right\rangle_0=\frac{C_\OO}{J|\tau|},
\qquad C_\OO=b^2=\frac{1}{2\sqrt\pi},
\label{eq:OOscaling}
\end{equation}
which corresponds to the scaling dimension
\begin{equation}
\Delta_\OO=\Delta_{\chi_L}+\Delta_{\chi_R}=\tfrac12 .
\label{eq:DeltaO}
\end{equation}
The coupling therefore has dimension $[g]=1-\Delta_\OO=\tfrac12>0$: the link is a \emph{relevant} perturbation of the decoupled conformal fixed point, which is why it matters at low energy even when it is weak.

The physical consequence is the retarded susceptibility $\chi^R_{\OO\OO}(t)=-i\theta(t)\langle[\OO(t),\OO(0)]\rangle$, which determines the linear response $\delta\langle\OO(t)\rangle=\int\dd t'\,\chi^R_{\OO\OO}(t-t')\,\delta g(t')$. A power law $1/|\tau|$ has a logarithmic Fourier transform, and causality fixes its phase:
\begin{equation}
\boxed{\;\frac1N\,\chi^R_{\OO\OO}(\omega)=-\frac{1}{\sqrt\pi\,J}\left[\ln\frac{J}{|\omega|}+i\,\frac\pi2\,\sgn\omega\right]+\chi^R_{\rm an}(\omega)\;}
\label{eq:chilog}
\end{equation}
where $\chi^R_{\rm an}$ is analytic at $\omega=0$ and collects ultraviolet and contact terms. The derivation is short. Equation~\eqref{eq:OOscaling} corresponds to a frequency-independent spectral weight $\sqrt\pi/J$ for $\omega>0$ at zero temperature, so that $\operatorname{Im}\chi^R=-\sqrt\pi/(2J)$ there. With our Fourier convention $\chi^R$ is analytic in the upper half plane and is therefore a function of $-i\omega$; analyticity then requires the logarithm and the phase to appear in the combination of Eq.~\eqref{eq:chilog}, with the same overall coefficient. A Kramers--Kronig evaluation of a constant spectral weight confirms $\operatorname{Re}\chi^R=-2\rho_0\ln(J/|\omega|)$ to within $2\%$ over $0.01\le\omega/J\le0.2$.

The logarithm has a clear physical meaning. The response to the link grows as the frequency is lowered, so the junction is sensitive to the slow part of the switching history, which a finite-time protocol only partly removes. It is also what makes the infrared of the junction the natural place to look for memory. At nonzero temperature the logarithm saturates at $\max(|\omega|,T)$.

\subsection{Numerical test of the conformal correlator}
\label{sec:euclidtest}

\begin{figure}[t]
\centering
\includegraphics[width=\linewidth]{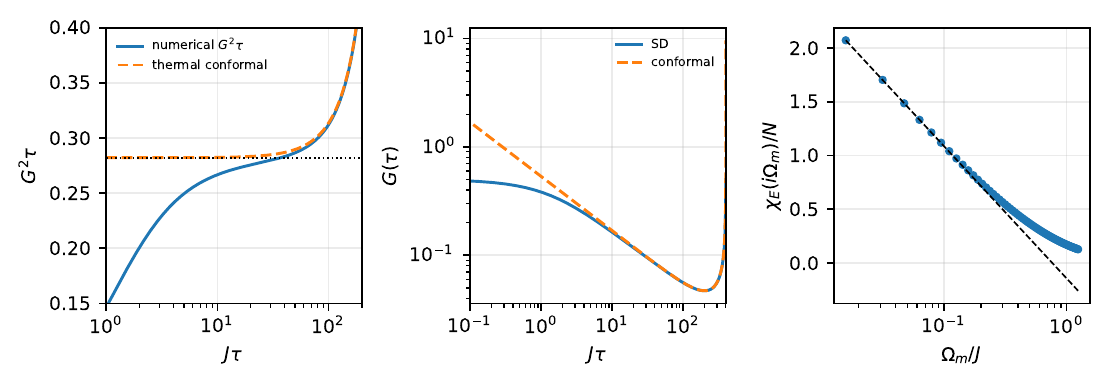}
\caption{Single-system Euclidean large-$N$ solution at $\beta J=400$. Left: $G^2\tau$ (solid), thermal conformal form (dashed) and $1/(2\sqrt\pi)$ (dotted). Middle: $G(\tau)$ against the thermal conformal form. Right: bosonic Matsubara transform of $G^2$ with its logarithmic fit.}
\label{fig:euclid}
\end{figure}

We tested Eqs.~\eqref{eq:Gc}--\eqref{eq:chilog} by solving the Euclidean large-$N$ equations for a single system at $\beta J=400$ on $2^{16}$ time points, with residual below $10^{-10}$ (Appendix~\ref{app:numerics}). Figure~\ref{fig:euclid} shows that $G(\tau)\to\tfrac12$ at short times ($G(0^+)=0.4995$), that $G^2\tau$ approaches the predicted $C_\OO=0.2821$ ($0.267$ at $J\tau=10$ and $0.280$ at $J\tau=30$), and that the logarithmic slope of $G$ for $8<J\tau<60$ is $-0.478$ against the conformal $-\tfrac12$. The bosonic Matsubara transform of $G^2$ has slope $-0.534$ against $\ln\Omega$ for $0.015<\Omega/J<0.06$, to be compared with the predicted $-1/\sqrt\pi=-0.564$. The $5\%$ difference is of the size expected from finite $\beta J$ and $1/(J\tau)$ corrections; we did not extrapolate.

\subsection{Work to first and second order in the drive}
\label{sec:workresponse}

The same susceptibility controls how much energy the switch deposits, and it determines how the work depends on the switching time. Let $\langle\OO\rangle_-$ denote the expectation value of the link operator just before the opening, in the weakly coupled state at $g_{\rm cl}$, and write $\delta g(t)=g(t)-g_{\rm cl}$. Linear response about that state gives $\langle\OO(t)\rangle=\langle\OO\rangle_-+\int\dd t'\,\chi^R_{\OO\OO}(t-t')\,\delta g(t')+O(\delta g^2)$, and the work of Eq.~\eqref{eq:work} becomes
\begin{equation}
W=\delta g\,\langle\OO\rangle_-+W_2,
\qquad
W_2=\int\dd t\,\dot g(t)\int\dd t'\,\chi^R_{\OO\OO}(t-t')\,\delta g(t').
\label{eq:Wsplit}
\end{equation}
The first term is a property of the state just before the opening. It does not depend on $\tau$, because $\int\dd t\,\dot g=\delta g$ for any protocol. Two remarks are in order. For the product initial state used in Sec.~\ref{sec:finiteN} one has $\langle\OO\rangle=0$ exactly at $t_i$: the state has definite parity on each side, so $\langle\chi^L_i\chi^R_i\rangle=\langle\chi^L_i\rangle\langle\chi^R_i\rangle=0$. The weak coupling $g_{\rm cl}$ then polarizes the state, producing a nonzero $\langle\OO\rangle_-=\chi_{\rm s}\,g_{\rm cl}$, where $\chi_{\rm s}$ is the static susceptibility, which is negative and extensive.

In frequency space the second-order term separates into an adiabatic and a dissipative part. The step-like profile of $g$ contributes a $\tau$-independent static piece $\tfrac12\chi^R_{\OO\OO}(0)\,\delta g^2$, which is negative because second-order perturbation theory lowers the energy, and a dissipative piece
\begin{equation}
W_{\rm diss}(\tau)=\frac{\delta g^2}{\pi}\int_0^\infty\!\dd\omega\;\frac{-\operatorname{Im}\chi^R_{\OO\OO}(\omega)}{\omega}\;S_\tau(\omega)^2\;\ge0,
\label{eq:Wdiss}
\end{equation}
where we used $|\widetilde g(\omega)|^2=\delta g^2S_\tau(\omega)^2/\omega^2$ (the transform of the profile itself carries a factor $1/\omega$ relative to that of $\dot g$). The filter $S_\tau^2$ removes the part of the integral above $\omega\sim1/\tau$, so $W_{\rm diss}$ \emph{decreases} with $\tau$ and $W_2$ becomes more negative. If $-\operatorname{Im}\chi^R_{\OO\OO}$ is constant, as in the conformal window of Eq.~\eqref{eq:chilog}, the integrand behaves as $1/\omega$ and the integral is logarithmic, cut off in the infrared at the scale $\mu$ where the susceptibility departs from its conformal form (the gap, the temperature or the finite-size level spacing):
\begin{equation}
W_{\rm diss}(\tau)\simeq\frac{N\,\delta g^2}{2\sqrt\pi\,J}\left[\ln\frac{1}{\mu\tau}+O(1)\right],
\qquad \frac1J\ll\tau\ll\frac1\mu .
\label{eq:Wlog}
\end{equation}
This is a prediction of the large-$N$ analysis: in the regime where the logarithm of Eq.~\eqref{eq:chilog} is operative, the work depends on the switching time through the same logarithm, $\dd W/\dd\ln\tau=-N\delta g^2/(2\sqrt\pi J)$. It makes the infrared structure of the link visible in an observable that can be measured, and it does not rely on any assumed quasiparticle map. We test the structure of Eq.~\eqref{eq:Wsplit} against exact dynamics in Sec.~\ref{sec:worktau}; the logarithm itself cannot be tested there, because the small systems have no window $1/J\ll\tau\ll1/\mu$.

%=====================================================================
\section{Where the interacting ladder enters}
\label{sec:ladder}

A natural worry is that the repeated scattering inside each SYK system dresses the response of Sec.~\ref{sec:response} through a four-time ladder, as it does for other correlators. Here we explain when it does and when it does not.

\subsection{Independent disorder means no cross ladder}
\label{sec:noladder}

The ladder kernel that dresses fluctuations of $G$ about the saddle is, for $q=4$,
\begin{equation}
K_c(1,2;3,4)=-3J^2\,G(1,3)\,G(2,4)\,G(3,4)^2,
\label{eq:ladder}
\end{equation}
up to a convention-dependent overall sign~\cite{MaldacenaStanford,PolchinskiRosenhaus}. Each rung is generated by the SYK interaction and connects two lines belonging to the \emph{same} system. Because the two disorder ensembles are independent, no rung connects an $L$ line to an $R$ line ($\Sigma_{LR}^{\rm SYK}=0$). Consequently:
\begin{itemize}[leftmargin=*]
\item the first-order cross-correlation $\delta G_{LR}$ is given exactly by Eq.~\eqref{eq:mainresponse}, with the exact propagators and no ladder resummation;
\item the junction bubble of Sec.~\ref{sec:susc} has no vertex correction at leading order;
\item the ladder acts on the \emph{diagonal} correlators $\delta G_{LL}$ and $\delta G_{RR}$, which first change at $O(g^2)$, where
\begin{equation}
\delta G_{LL}=(1-K_C)^{-1}\circ\delta G^{(2)}_{LL}.
\label{eq:dressed}
\end{equation}
\end{itemize}
A cross ladder would require correlated disorder between the two systems, as in the Maldacena--Qi construction~\cite{MaldacenaQi}, in which the infrared physics is also different (Sec.~\ref{sec:discussion}). Equation~\eqref{eq:dressed} is the correct starting point for any attempt to derive an effective quasiparticle map, which must be built from the second-order response.

\subsection{Projection and a separability diagnostic}
\label{sec:projection}

Suppose a projection $\Pc_{\rm qp}$ maps the interacting two-time response to a linear quasiparticle amplitude $\beta_{\rm eff}=\Pc_{\rm qp}\,\delta G$. A tractable question is whether $\beta_{\rm eff}$ has the factorized form
\begin{equation}
\beta(\omega,\nu)=f(\omega+\nu)\,h(\omega)\,k(\nu),
\label{eq:factor}
\end{equation}
which is what a source with a sum-frequency dependence and separately dressed legs would produce. If it does, then
\begin{equation}
\partial_\omega\partial_\nu\ln\beta=\frac{\dd^2}{\dd s^2}\ln f(s)\Big|_{s=\omega+\nu}.
\label{eq:separability}
\end{equation}
A dependence on the relative frequency $\omega-\nu$ that cannot be absorbed into $h$ and $k$ would violate this relation. The test applies to numerical or analytic solutions of Eq.~\eqref{eq:dressed}. Note that the first-order response, Eq.~\eqref{eq:twosource}, has the form~\eqref{eq:factor} \emph{identically}, since it is a product of a function of $\omega+\nu$ and functions of $\omega$ and $\nu$; the diagnostic can therefore detect a violation only at higher order. We do not attempt that calculation here.

%=====================================================================
\section{A conditional kernel and its infrared norm}
\label{sec:benchmark}

In a free Majorana shutter, a sudden rotation produces a Bogoliubov kernel $\beta_{\rm free}\propto\sin\theta/(\omega+\nu)$, whose Hilbert--Schmidt (HS) norm grows logarithmically with the infrared cutoff~\cite{Vahedi2026}. It is natural to ask how an interacting version of this kernel would behave. We treat this as a controlled benchmark, not as a result about SYK.

\subsection{The benchmark and its norm}
\label{sec:hs}

Consider the conditional ansatz
\begin{equation}
\beta_{\rm ans}(\omega,\nu)=A\,\frac{e^{i(\omega+\nu)t_0}\,S_\tau(\omega+\nu)}{(\omega+\nu)\,(\omega\nu)^{1/4}},
\qquad\omega,\nu>0 .
\label{eq:ansatz}
\end{equation}
The factor $(\omega\nu)^{-1/4}$ is a spectral \emph{amplitude}: it is the square root of a density of states $\rho\sim\omega^{-1/2}$, as would arise if each leg of the kernel carried the square root of the conformal spectral weight. This is a hypothesis about the projection, not a derived result. In the sudden regime $\tau(\omega+\nu)\ll1$ one has $|\beta_{\rm ans}|^2=|A|^2/[\sqrt{\omega\nu}\,(\omega+\nu)^2]$. With an infrared cutoff $\mu$ and conformal upper scale $\omega_\ast$ the HS norm is $|A|^2I(\mu,\omega_\ast)$ with
\begin{equation}
I(\mu,\omega_\ast)=\int_\mu^{\omega_\ast}\!\!\dd\omega\int_\mu^{\omega_\ast}\!\!\dd\nu\;\frac{1}{\sqrt{\omega\nu}\,(\omega+\nu)^2}
=\frac{C}{\mu}+O(\omega_\ast^{-1}),
\qquad
C=\frac\pi2-1 .
\label{eq:HS}
\end{equation}
Scaling out $\mu$ shows the $1/\mu$ growth, and the constant is evaluated in Appendix~\ref{app:hs}.

\subsection{Numerical check and comparison with other kernels}
\label{sec:hsnum}

\begin{figure}[t]
\centering
\includegraphics[width=0.92\linewidth]{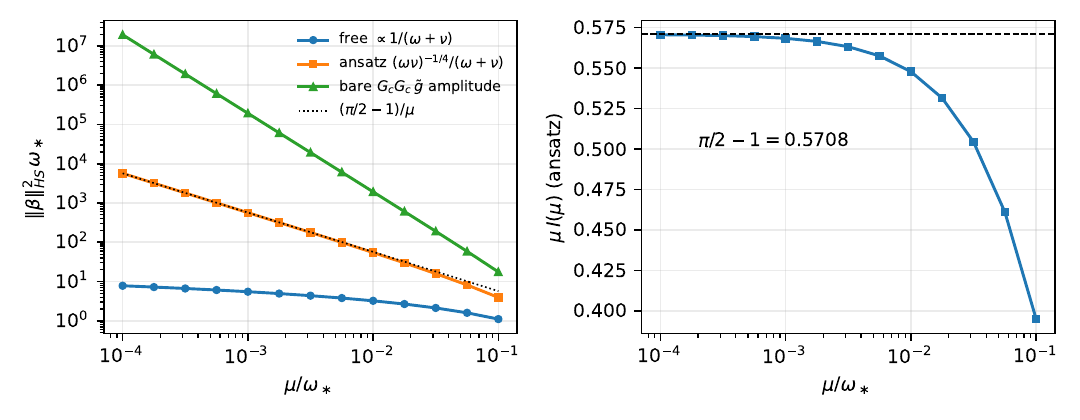}
\caption{Infrared behavior of the Hilbert--Schmidt norm for three kernels with the same sudden-limit source: free (logarithmic), benchmark ($1/\mu$) and bare conformal propagators ($1/\mu^2$), together with the convergence of the coefficient $C$ to $\pi/2-1$.}
\label{fig:hs}
\end{figure}

We checked the coefficient by direct quadrature. At $\omega_\ast/\mu=10^{4}$, $\mu I$ equals $0.5705$ at $\mu=10^{-4}$ and $0.5683$ at $\mu=10^{-3}$, against $\pi/2-1=0.5708$; the difference is the expected $O(\mu/\omega_\ast)$ tail. To place the benchmark in context we compared the infrared behavior of three kernels with the same sudden-limit source (Fig.~\ref{fig:hs}). The free kernel $1/(\omega+\nu)$ grows as $\ln(1/\mu)$ (fitted coefficient $0.998$); the benchmark grows as $\mu^{-1.005}$; and the literal product of two conformal propagators and the source, $G_c(\omega)\widetilde g(\omega+\nu)G_c(\nu)$ with $|G_c|\sim\omega^{-1/2}$, grows as $\mu^{-2.000}$. The benchmark sits between the free and bare cases because it carries a spectral amplitude instead of a propagator on each leg. The result is correct for the assumed kernel; its physical status depends on whether the ladder-dressed projection actually produces Eq.~\eqref{eq:ansatz}, which remains open.

%=====================================================================
\section{Infrared scales and the crossover}
\label{sec:scales}

A relevant coupling cannot be treated perturbatively down to arbitrarily low energy. Here we estimate where the decoupled description fails and test the estimate against the coupled equilibrium problem.

\subsection{Dimensional estimate}
\label{sec:dimestimate}

Since $[g]=\tfrac12$, the natural dimensionless coupling at an energy scale $E$ is $\lambda(E)\sim g/\sqrt{JE}$, which grows as $\lambda\sim E^{-1/2}$ toward the infrared. The absence of an $O(\lambda^2)$ term in its flow follows from the symmetry $\chi^R\to-\chi^R$ of $H_R$, which maps $g\to-g$. The decoupled conformal expansion ceases to be controlled when $\lambda(E_g)\sim1$:
\begin{equation}
E_g\sim\frac{g^2}{J},
\label{eq:Eg}
\end{equation}
up to a convention-dependent number of order unity. This is a scaling estimate, not a one-loop determination of a universal coefficient, and a divergent perturbative susceptibility signals the breakdown of the decoupled saddle and not an observable infinity. The natural crossover variables are $\omega/E_g$, $T/E_g$ and $E_g\tau$, together with the finite-size scale $E_{\rm IR}(N)/E_g$.

\subsection{Numerical test: the coupled equilibrium problem}
\label{sec:gaptest}

\begin{figure}[t]
\centering
\includegraphics[width=0.92\linewidth]{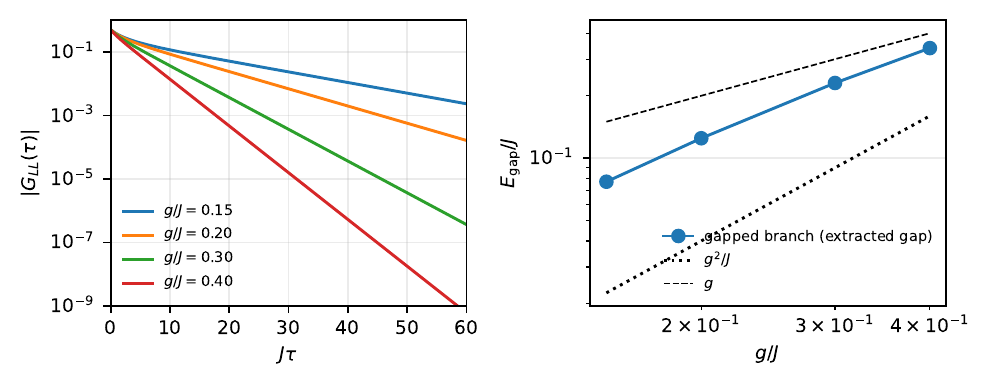}
\caption{Coupled equilibrium large-$N$ solution with independent disorder. Left: $|G_{LL}(\tau)|$ for several couplings. Right: extracted decay rate $E_{\rm gap}$ against $g/J$, with $g^2/J$ (dotted) and $g$ (dashed) for reference.}
\label{fig:gap}
\end{figure}

With independent disorder the equilibrium coupled problem closes on a single diagonal self-energy. The key observation is that the link is a chemical potential of magnitude $g$ for the complex fermion $c=(\chi^L+i\chi^R)/\sqrt2$, so that in Matsubara frequency, with $A(i\omega)=-i\omega-\Sigma(i\omega)$ and $\Sigma(\tau)=J^2G_{LL}(\tau)^3$,
\begin{equation}
G_{LL}(i\omega)=\frac12\left[\frac{1}{A+g}+\frac{1}{A-g}\right]=\frac{A}{A^2-g^2}.
\label{eq:GLLeq}
\end{equation}
We solved this self-consistent problem at low temperature (residual below $10^{-9}$) and extracted the decay rate $E$ of $G_{LL}(\tau)$ at long times (Fig.~\ref{fig:gap}). We find $E/J=0.077,\,0.125,\,0.230$ at $g/J=0.15,\,0.2,\,0.3$ ($\beta J=1000,\,800,\,400$). The local exponent in this window is about $1.5$--$1.6$, and the ratio $E/g^2$ rises from $2.6$ to $3.4$ as $g$ decreases. The data therefore do not exhibit $E\propto g^2/J$. For $g/J\le0.12$ the available $\beta J\le1000$ does not reach the regime $\beta E\gg1$, so the $g\to0$ asymptotics is unresolved. The value at $g/J=0.4$ is less reliable because the signal reaches the numerical floor, and is shown only for completeness.

We conclude that Eq.~\eqref{eq:Eg} should be read as the scale at which the decoupled expansion fails, not as a measured gap: in the accessible window the actual gap is larger and grows more slowly with $g$. Whether the true asymptotic scaling is recovered at smaller $g$ cannot be decided from these data.

\subsection{Temperature and finite size}
\label{sec:finiteT}

At nonzero temperature the conformal correlator follows from the usual thermal map, $\tau\to(\beta/\pi)\sin(\pi\tau/\beta)$, so the zero-temperature logarithm of Eq.~\eqref{eq:chilog} is cut off at $\max(|\omega|,T)$ within the conformal window and above $E_g$. At finite $N$ the spectrum is discrete, and the lowest resolvable scale $E_{\rm IR}(N)$ depends on the disorder realization and the observable; it should not be replaced by a universal $J/N$ estimate. The order of limits matters: one must say how $N\to\infty$, $T\to0$ and $g\to0$ are taken relative to one another. In particular, the logarithm is a statement about a window of scales, not an infinite quantity in a finite Hilbert space.

%=====================================================================
\section{Exact dynamics of finite systems}
\label{sec:finiteN}

The large-$N$ analysis gives analytic control but only for the leading response, and it cannot follow strong driving. For small $N$ the opposite is true: the full unitary evolution can be computed exactly, including entanglement, parity transfer and scrambling, but there is no infrared window. The two descriptions therefore complement each other. In this section we follow the exact dynamics for $N=6$ to $12$ Majoranas per system, using disorder averages with error bars, and we ask three questions. How much of an initial excitation moves across the link, and how entangled do the two systems become? How do these depend on the switching time? And, at equal injected work, does the outcome still remember how the link was opened?

\subsection{Setup and initial state}
\label{sec:setup}

We represent the Majoranas by Jordan--Wigner matrices $\gamma_i$ with $\{\gamma_i,\gamma_j\}=2\delta_{ij}$ and $\chi_i=\gamma_i/\sqrt2$. Two systems with $N$ Majoranas each contain $2N$ Majoranas, i.e.\ $N$ complex fermionic modes, so $\dim\mathcal H=2^N$ and $\dim\mathcal H_L=\dim\mathcal H_R=2^{N/2}$ ($256$ and $16$ for $N=8$; $4096$ and $64$ for $N=12$). The initial state is a localized odd-parity excitation of the left system on top of the product of the two uncoupled ground states,
\begin{equation}
|\Psi_i\rangle=\frac{c_L^\dagger\,|\Omega_L\Omega_R\rangle}{\sqrt{\langle\Omega_L\Omega_R|\,c_L c_L^\dagger\,|\Omega_L\Omega_R\rangle}},
\qquad
c_L^\dagger=\tfrac12\left(\gamma_1^L-i\gamma_2^L\right),
\label{eq:initstate}
\end{equation}
where the denominator normalizes the state. A subtle point is the choice of $|\Omega_a\rangle$. The single-system spectra fall into symmetry classes that depend on $N$ modulo $8$~\cite{FuSachdev,GarciaGarciaVerbaarschot}: for $N=6,10$ the lowest levels are degenerate between the two parity sectors, and for $N=12$ they are Kramers-degenerate. An arbitrary vector from the ground-state manifold may then be parity-mixed or entangled between the two systems. We therefore take $|\Omega_L\rangle$ and $|\Omega_R\rangle$ to be the lowest even-parity eigenstate of each system \emph{separately}. With this prescription the initial entropy vanishes to $4\times10^{-16}$ at every size, and the initial state is a clean parity excitation.

The state evolves under $i\,\partial_t|\Psi\rangle=H(t)|\Psi\rangle$ from $Jt_i=-6$ to $Jt_f=14$, with $g_{\rm cl}/J=0.05$, $g_{\rm op}/J=0.70$ and $Jt_0=2$, so the system evolves at weak coupling for $Jt_0-Jt_i=8$ before the opening; some entanglement already builds up during this period (Fig.~\ref{fig:ed}). We use a fourth-order Runge--Kutta integrator with step $0.02$ (propagators: $0.04$, in the two total-parity blocks). Error bars are one standard error of the mean over independent disorder realizations: $24$ for the three-protocol study at $N=8$, $40,40,40,13$ for the size scan at $N=6,8,10,12$, $6$ for the OTOC, and $40$ ($N=8$) and $13$ ($N=12$) for the equal-work test. The $N=12$ statistics are modest, and the error bars should be read accordingly. These are finite-size diagnostics, not estimates of a thermodynamic limit.

\subsection{Consistency checks}
\label{sec:checks}

The integration is accurate. The norm drift is at most $8.3\times10^{-9}$ for $N=8$ and $3.8\times10^{-8}$ for $N=12$; the drift of total parity is at most $1.7\times10^{-8}$; the propagator is unitary to $5.8\times10^{-7}$; and the work integral of Eq.~\eqref{eq:work} agrees with the directly measured energy change to better than $2.2\times10^{-6}$ for every realization and protocol. This confirms the exact work identity numerically.

\subsection{Parity transfer and entanglement}
\label{sec:parity}

\begin{figure}[t]
\centering
\includegraphics[width=\linewidth]{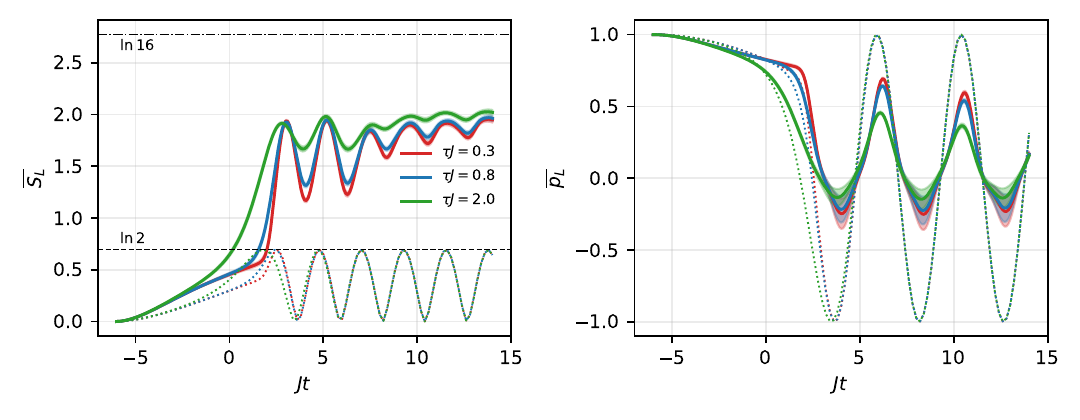}
\caption{Disorder-averaged entropy $\overline{S_L}$ (left) and normalized left parity $\overline{p_L}$ (right) for $\tau J=0.3,0.8,2.0$ at $N=8$ (bands: $\pm1$ SEM over $24$ realizations). Dotted lines: free-fermion reference. Horizontal lines: $\ln2$ and $\ln16$.}
\label{fig:ed}
\end{figure}

The first question is how much of the left-parity excitation reaches the right system and how entangled the two systems become. Figure~\ref{fig:ed} shows the disorder-averaged normalized parity $\overline{p_L(t)}$ and entropy $\overline{S_L(t)}$ for the three switching times, together with a free reference ($J=0$) in which every flavor pair rotates independently.

In the free reference the parity oscillates between $\pm1$ with the link rotation, and the entropy is bounded by $\ln2=0.693$ (the computed maximum is $0.6931$), because a single fermion delocalized between the two systems can share at most one bit. The interacting system behaves very differently. The parity oscillations are strongly damped, and the ensemble-averaged entropy rises to $1.95$, $1.97$ and $2.03$ for $\tau J=0.3,0.8,2.0$ at $N=8$, about $2.8$ times the free ceiling and $70$--$73\%$ of the Hilbert-space bound $4\ln2=2.77$. An entropy above $\ln2$ is not a failure of the large-$N$ picture: it shows that the interacting state has left the one-particle subspace in which a quadratic description would hold.

\subsection{Dependence on the switching time}
\label{sec:tau}

Table~\ref{tab:tau8} gives the extrema for $N=8$. The maximum entropy increases weakly with $\tau$, the parity excursion decreases in magnitude with $\tau$, and $|W|$ increases with $\tau$. The sample-to-sample scatter of the minimum parity is large at this size (standard deviation $0.2$--$0.3$), so differences between a handful of realizations should not be read as trends.

\begin{table}[t]
\centering
\begin{tabular}{cccc}
\toprule
$\tau J$ & $\overline{\max_t S_L}$ & $\overline{\min_t p_L}$ & $\overline W/J$\\
\midrule
$0.3$ & $2.039\pm0.012$ & $-0.41\pm0.06$ & $-0.233\pm0.012$\\
$0.8$ & $2.047\pm0.012$ & $-0.37\pm0.06$ & $-0.315\pm0.015$\\
$2.0$ & $2.090\pm0.014$ & $-0.24\pm0.04$ & $-0.475\pm0.023$\\
\bottomrule
\end{tabular}
\caption{Disorder-averaged extrema and total work at $N=8$ (mean $\pm$ SEM, $24$ realizations).}
\label{tab:tau8}
\end{table}

\subsection{Work}
\label{sec:work_ed}

The total work is negative for all three protocols, and its magnitude \emph{increases} with $\tau$ (Fig.~\ref{fig:work}, left and middle). The sign has a simple origin. The weak coupling $g_{\rm cl}$ polarizes the state before the opening so that $\langle\OO\rangle_-<0$, and in a sudden opening the work is $\delta g\,\langle\OO\rangle_-$ (Sec.~\ref{sec:workresponse}). For $\delta g=0.65J$ at $N=8$ this term accounts for $92\%$ of the work at $\tau J=0.3$ but only $44\%$ at $\tau J=2.0$: the growth of $|W|$ with $\tau$ comes from the second-order part, which becomes more negative as the opening slows. This is the opposite of a $1/\tau$ law. Section~\ref{sec:worktau} analyzes this decomposition quantitatively.

\subsection{Cross-system scrambling}
\label{sec:otoc}

\begin{figure}[t]
\centering
\includegraphics[width=\linewidth]{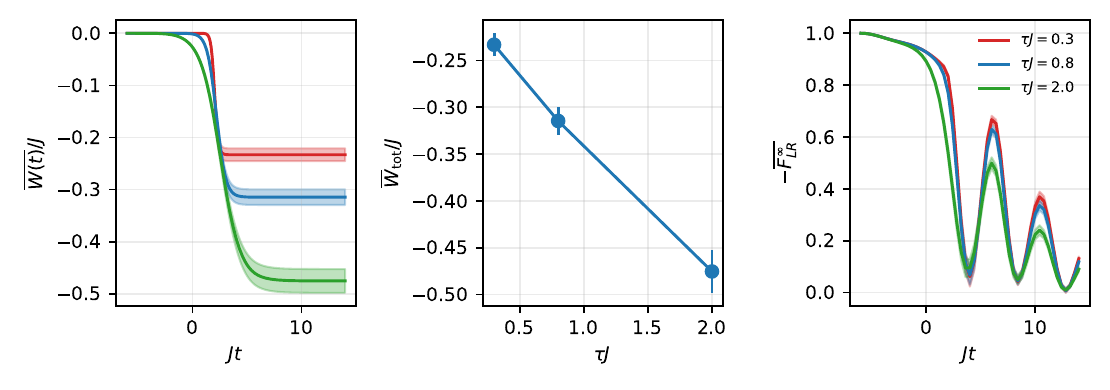}
\caption{Left: ensemble-averaged cumulative work $\overline{W(t)}$. Middle: total work against $\tau J$. Right: $-F_{LR}(t)$ for the three protocols ($6$ realizations). Bands: $\pm1$ SEM.}
\label{fig:work}
\end{figure}

The OTOC of Eq.~\eqref{eq:OTOC} (Fig.~\ref{fig:work}, right) starts at $1$, drops to values near $0$ at $Jt\simeq4$--$5$ for all three protocols, partially revives (to about $0.6$ for the fastest and $0.45$ for the slowest switching at $Jt\simeq6$), returns near zero and shows a smaller bump at late times. We do not fit a Lyapunov exponent: with eight Majoranas per system there is no exponential regime. The first revival follows the opening by roughly $\pi/g_{\rm op}\approx4.5$, the period of the free link rotation, which suggests that it is driven by the coherent rotation induced by the link rather than by a many-body recurrence. The timing is the same for all three protocols (Sec.~\ref{sec:otocmemory}), which is consistent with this picture but does not prove it; we did not vary $g_{\rm op}$. Six realizations are used here; Sec.~\ref{sec:otocmemory} repeats the calculation with $24$.

\subsection{Dependence on system size}
\label{sec:size}

\begin{figure}[t]
\centering
\includegraphics[width=\linewidth]{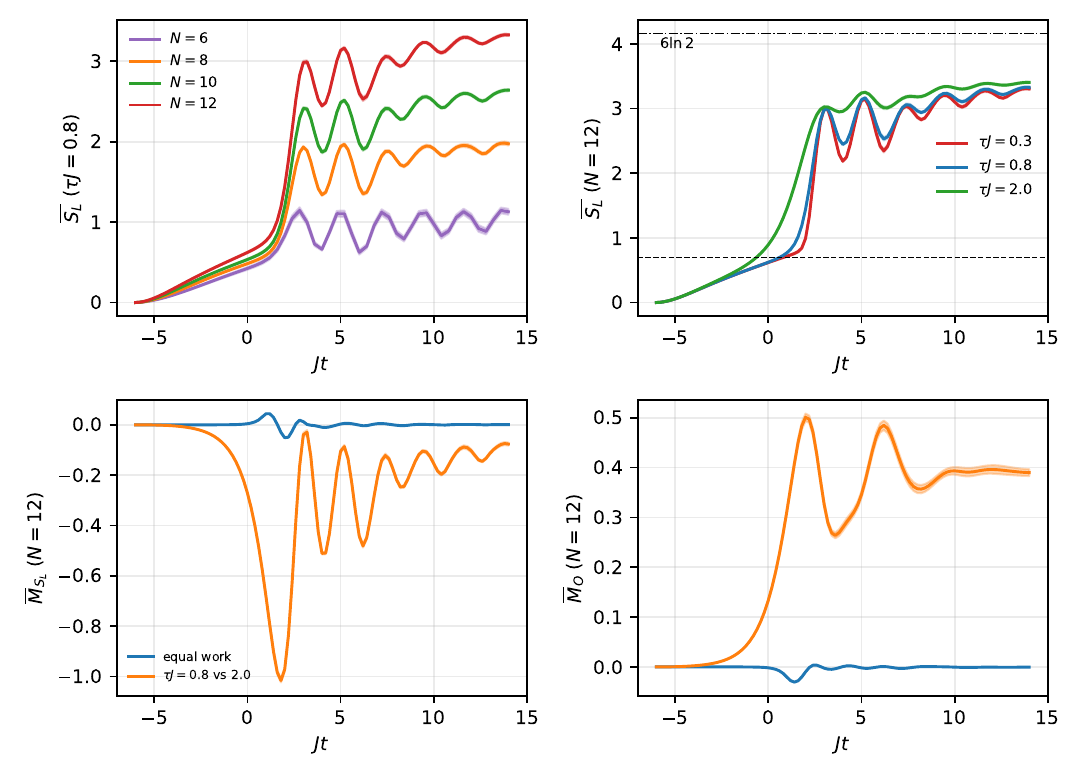}
\caption{Size dependence. Top left: $\overline{S_L(t)}$ at $\tau J=0.8$ for $N=6,8,10,12$. Top right: $N=12$ for the three switching times, with $\ln2$ and $6\ln2$. Bottom: memory functionals $\overline{\mathcal M}_{S_L}$ and $\overline{\mathcal M}_{\OO}$ at $N=12$ for equal-work protocols (blue; tanh against ramp) and for unmatched tanh protocols $\tau J=0.8$ against $2.0$ (orange). Bands: $\pm1$ SEM.}
\label{fig:n12}
\end{figure}

\begin{figure}[t]
\centering
\includegraphics[width=\linewidth]{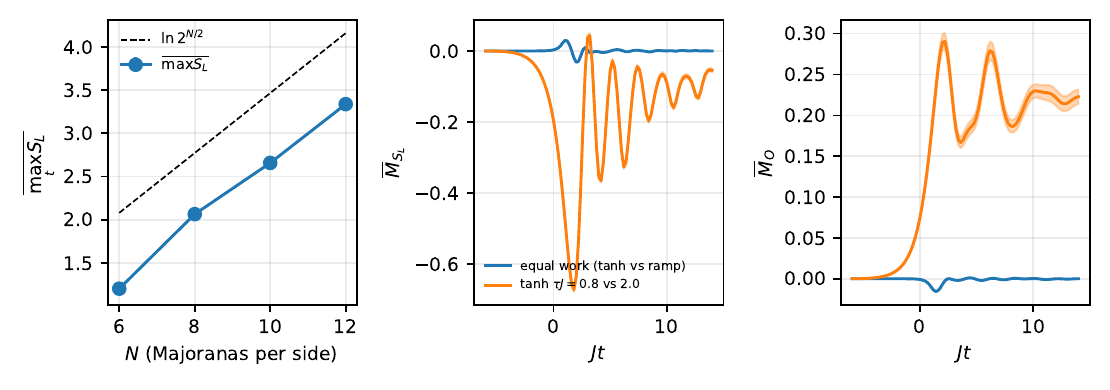}
\caption{Left: maximum entropy against $N$ with the bound $\tfrac N2\ln2$. Middle and right: memory functionals $\overline{\mathcal M}_{S_L}$ and $\overline{\mathcal M}_{\OO}$ at $N=8$ ($40$ realizations) for equal-work protocols (blue; tanh against ramp) and for unmatched $\tau J=0.8$ against $2.0$ (orange); bands $\pm1$ SEM.}
\label{fig:scale}
\end{figure}

To see whether these effects survive at larger size we repeated the calculation for $N=6,8,10,12$ ($24$ Majoranas in total and Hilbert-space dimension $4096$ at $N=12$). The maximum entropy at $\tau J=0.8$ is $1.202\pm0.037$, $2.064\pm0.011$, $2.657\pm0.012$ and $3.337\pm0.012$, i.e.\ $58\%$, $74\%$, $77\%$ and $80\%$ of the bounds $\tfrac N2\ln2$ (Fig.~\ref{fig:n12}, top left, and Fig.~\ref{fig:scale}, left). From $N=8$ to $12$ the entropy grows by $0.32$ per added Majorana, compared with the maximum $\tfrac12\ln2=0.347$. This is compatible with volume-law growth with a Page-like deficit, but four small sizes do not prove it, and $N=6$ belongs to a different symmetry class of the single-system spectrum. At $N=12$ the entropy is about $4.8$ times the free ceiling.

Table~\ref{tab:tau12} gives the $N=12$ analogue of Table~\ref{tab:tau8}. The trends of Sec.~\ref{sec:tau} persist, and two size effects stand out. The work is approximately extensive: at $\tau J=0.8$ it grows from $-0.315$ to $-0.477$, a ratio of $1.5$ for a ratio of $1.5$ in $N$. And parity transfer becomes self-averaging: the standard deviation of $\min p_L$ at $\tau J=0.8$ falls from about $0.26$ at $N=8$ to about $0.03$ at $N=12$, so that the large disorder fluctuations of the smaller system give way to a well-defined mean behavior.

\begin{table}[t]
\centering
\begin{tabular}{cccc}
\toprule
$\tau J$ & $\overline{\max_t S_L}$ & $\overline{\min_t p_L}$ & $\overline W/J$\\
\midrule
$0.3$ & $3.315\pm0.014$ & $-0.365\pm0.009$ & $-0.325\pm0.007$\\
$0.8$ & $3.337\pm0.012$ & $-0.294\pm0.009$ & $-0.477\pm0.009$\\
$2.0$ & $3.414\pm0.008$ & $-0.136\pm0.006$ & $-0.751\pm0.012$\\
\bottomrule
\end{tabular}
\caption{As Table~\ref{tab:tau8} for $N=12$ (mean $\pm$ SEM, $13$ realizations).}
\label{tab:tau12}
\end{table}

\subsection{Memory at equal work}
\label{sec:memory}

The question that motivates this paper is whether the late-time state remembers how the coupling was opened, beyond the energy deposited. Comparing two tanh protocols with different $\tau$ is not a clean test, since they inject different work (Table~\ref{tab:tau8}). We therefore compare the $\tau J=0.8$ tanh protocol with a linear ramp, centered at $t_0$, with the same $g_{\rm cl}$ and $g_{\rm op}$, whose duration is tuned \emph{per realization} so that the two protocols deposit the same energy. We match on the exact total energy change of Eq.~\eqref{eq:work}; matching by quadrature of $\dot g\langle\OO\rangle$ is inaccurate for a ramp, whose derivative is discontinuous. The mean ramp duration is $JT_r=2.36$ at $N=8$ and $2.35$ at $N=12$, and the maximum residual work mismatch is $3.5\times10^{-5}$ in both cases.

Table~\ref{tab:mem} gives the memory functionals of Eq.~\eqref{eq:memory}, averaged over $8\le Jt\le14$, for the entropy and for $\langle\OO\rangle$. For comparison we also give the same quantities for two tanh protocols with $\tau J=0.8$ and $2.0$ without work matching. Matching the work removes more than $99.7\%$ of the protocol dependence of the late-time entropy and more than $99.9\%$ of that of $\langle\OO\rangle$ at both sizes. The entropy difference is within about two standard errors of zero. The $\langle\OO\rangle$ difference is formally nonzero ($5$--$6\sigma$) but is about $10^{-3}$ of the unmatched effect, with the same relative size at $N=8$ and $12$.

\begin{table}[t]
\centering
\footnotesize
\setlength{\tabcolsep}{4pt}
\begin{tabular}{lcccc}
\toprule
 & \multicolumn{2}{c}{$N=8$ ($40$ samples)} & \multicolumn{2}{c}{$N=12$ ($13$ samples)}\\
 & $\Delta S_L$ & $\Delta\langle\OO\rangle$ & $\Delta S_L$ & $\Delta\langle\OO\rangle$\\
\midrule
equal work (tanh vs ramp) & $0.0002\pm0.0001$ & $-0.00016\pm0.00003$ & $0.0003\pm0.0001$ & $-0.00031\pm0.00005$\\
unmatched ($\tau J=0.8$ vs $2.0$) & $-0.107\pm0.004$ & $0.217\pm0.009$ & $-0.138\pm0.004$ & $0.387\pm0.007$\\
\bottomrule
\end{tabular}
\caption{Memory functionals at late times ($8\le Jt\le14$), mean $\pm$ SEM.}
\label{tab:mem}
\end{table}

We therefore see no evidence of protocol memory beyond energy deposition within this protocol family. The small residual in $\langle\OO\rangle$ is not necessarily memory: matching the mean energy does not match its variance or higher moments, and the two protocol shapes differ in those. The null result should also be read against the absence of an infrared window at these sizes, where the level spacing is of order $J$; it does not test the logarithmic memory of the large-$N$ analysis. Natural extensions are an initial state localized on the right or unexcited, the pure-state OTOC, the equal-work OTOC at larger sizes, protocols matched in the energy variance, and more realizations at $N=12$.

\subsection{Scrambling at equal work}
\label{sec:otocmemory}

The revival height of the OTOC depends on the switching time (Fig.~\ref{fig:work}), so the same question arises as for the entropy: is this protocol dependence anything more than the effect of different injected energy? We tested it at $N=8$ with the work-matched protocols of Sec.~\ref{sec:memory}, for the same $24$ disorder realizations and the same per-realization ramp durations, evolving the full propagator for the tanh protocols with $\tau J=0.8$ and $2.0$ and for the matched ramp.

\begin{figure}[t]
\centering
\includegraphics[width=\linewidth]{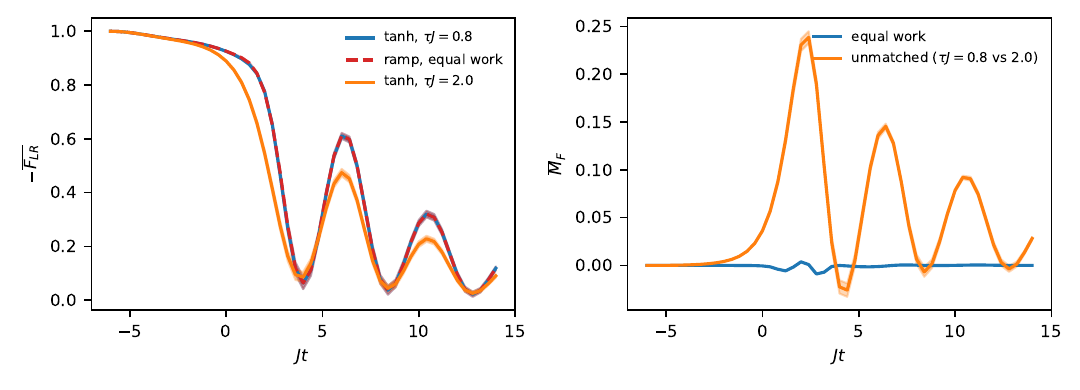}
\caption{Cross-system OTOC at equal work ($N=8$, $24$ realizations; bands $\pm1$ SEM). Left: $-\overline{F_{LR}}$ for the tanh protocols with $\tau J=0.8$ and $2.0$ and for the ramp matched in work to the former. Right: the corresponding memory functional of the OTOC, for equal work (blue) and unmatched (orange).}
\label{fig:otoceq}
\end{figure}

The result mirrors the entropy (Fig.~\ref{fig:otoceq}). The height of the first revival, reached at $Jt=6.02\pm0.02$ for all three protocols, is $0.611\pm0.011$ for the $\tau J=0.8$ tanh protocol, $0.612\pm0.011$ for the work-matched ramp and $0.474\pm0.014$ for the $\tau J=2.0$ tanh protocol. The difference between the matched protocols is $-0.0012\pm0.0001$, compared with $0.136\pm0.003$ without matching. At late times ($8\le Jt\le14$) the matched difference is $0.00013\pm0.00005$, against $0.033\pm0.002$ unmatched, and the root-mean-square difference of the ensemble means over $Jt\ge2$ falls from $0.094$ to $0.0022$, a reduction of about $98\%$. The protocol dependence of the scrambling diagnostic is therefore also controlled by the injected energy, to within a residual of order one percent of the unmatched effect. We did not repeat this at $N=10$ or $12$, where the full propagator is expensive, and the caveats about unmatched energy variance and the absence of an infrared window apply here too.

\subsection{How the work depends on the switching time}
\label{sec:worktau}

\begin{figure}[t]
\centering
\includegraphics[width=\linewidth]{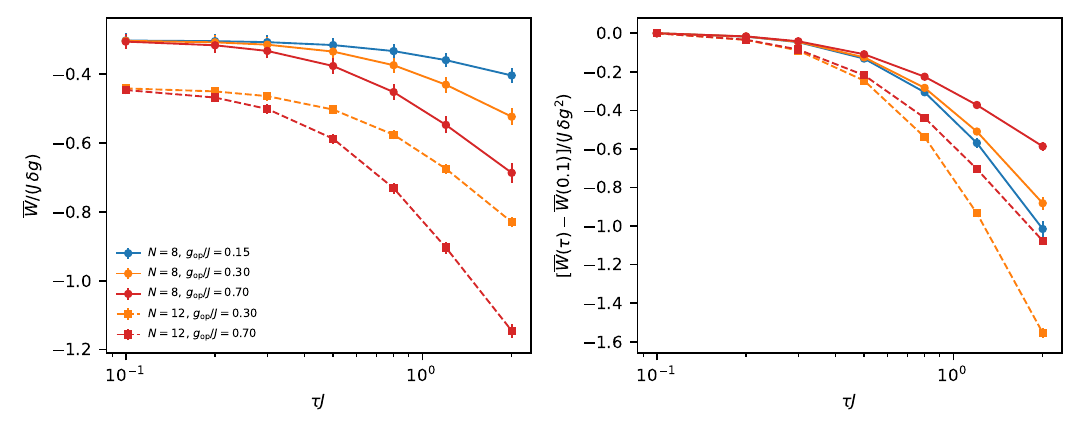}
\caption{Dependence of the work on the switching time. Left: $\overline W/(J\,\delta g)$ against $\tau J$ for several final couplings $g_{\rm op}$ and for $N=8$ (solid, $24$ realizations) and $N=12$ (dashed, $10$ realizations). For each $N$ the curves at different $g_{\rm op}$ collapse onto $\langle\OO\rangle_-$ at small $\tau$. Right: the $\tau$-dependent part, $[\overline W(\tau)-\overline W(0.1J^{-1})]/(J\,\delta g^2)$. Bars: $\pm1$ SEM.}
\label{fig:worktau}
\end{figure}

The decomposition~\eqref{eq:Wsplit} makes two testable statements about the exact finite-size data. The first-order term $\delta g\,\langle\OO\rangle_-$ should be independent of $\tau$, linear in $\delta g$ and extensive in $N$. The remainder is of second order in $\delta g$ and should depend on $\tau$. We tested both by computing $W$ for seven switching times, $\tau J=0.1$ to $2.0$, and for several final couplings, with $24$ realizations at $N=8$ and $10$ at $N=12$ (Fig.~\ref{fig:worktau}).

\emph{First-order part.} In the sudden limit, $\tau J=0.1$, the ratio $W/\delta g$ is $-0.302$, $-0.303$ and $-0.305$ for $\delta g/J=0.10$, $0.25$ and $0.65$ at $N=8$, so $W$ is linear in $\delta g$ there and $\langle\OO\rangle_-\simeq-0.30$. At $N=12$ the corresponding values are $-0.44$ and $-0.45$ for $\delta g/J=0.25$ and $0.65$. The ratio $1.46$ between the two sizes is close to $1.5$, so $\langle\OO\rangle_-\simeq-0.037N$ is extensive. The initial value $\langle\OO\rangle$ at $t_i$ vanishes identically, as argued above.

\emph{Second-order part.} The $\tau$-dependent part $[W(\tau)-W(0.1J^{-1})]/\delta g^2$ is shown in the right panel of Fig.~\ref{fig:worktau}. At weak coupling it is approximately independent of $\delta g$ (it equals $-1.01$ at $\tau J=2.0$ for $\delta g/J=0.10$ and $-0.88$ for $0.25$) and it bends away at $\delta g/J=0.65$ ($-0.59$), where the second-order expansion is no longer adequate. It grows monotonically in magnitude with $\tau$ across the accessible range and shows no saturation by $\tau J=2$; the adiabatic limit is not reached. There is no $1/\tau$ behavior.

\emph{Comparison with the conformal prediction.} The logarithmic law~\eqref{eq:Wlog} predicts a slope $\dd W/\dd\ln\tau=-N\delta g^2/(2\sqrt\pi J)$, i.e.\ $-0.95$ at $N=8$ and $-1.4$ at $N=12$ for $\delta g/J=0.65$. The observed slopes between $\tau J=0.5$ and $2.0$ are $-0.15$ and $-0.26$, smaller by a factor of about $6$. This is not a failure of the prediction, which concerns the conformal window $1/J\ll\tau\ll1/\mu$, but it shows that the accessible sizes and switching times are far from it: at $N=8$ to $12$ the frequencies $\omega\sim1/\tau$ probed here are comparable to $J$ and to the level spacing, where the spectral weight of $\OO$ is not constant. We therefore make no quantitative claim of agreement; what the data do support is the structure of Eq.~\eqref{eq:Wsplit}, the sign, the extensivity of the first-order term, and the second-order character of the $\tau$ dependence.

\subsection{Equal-work protocols that differ at low or at high frequency}
\label{sec:tuned}

The null result of Sec.~\ref{sec:memory} compares two protocols that differ at all frequencies. Since the question is about \emph{infrared} memory, it is natural to ask whether protocols whose difference is concentrated at low frequency leave a larger residual at equal work than protocols whose difference is at high frequency. We tested this with a controlled pair of comparisons.

\begin{figure}[t]
\centering
\includegraphics[width=\linewidth]{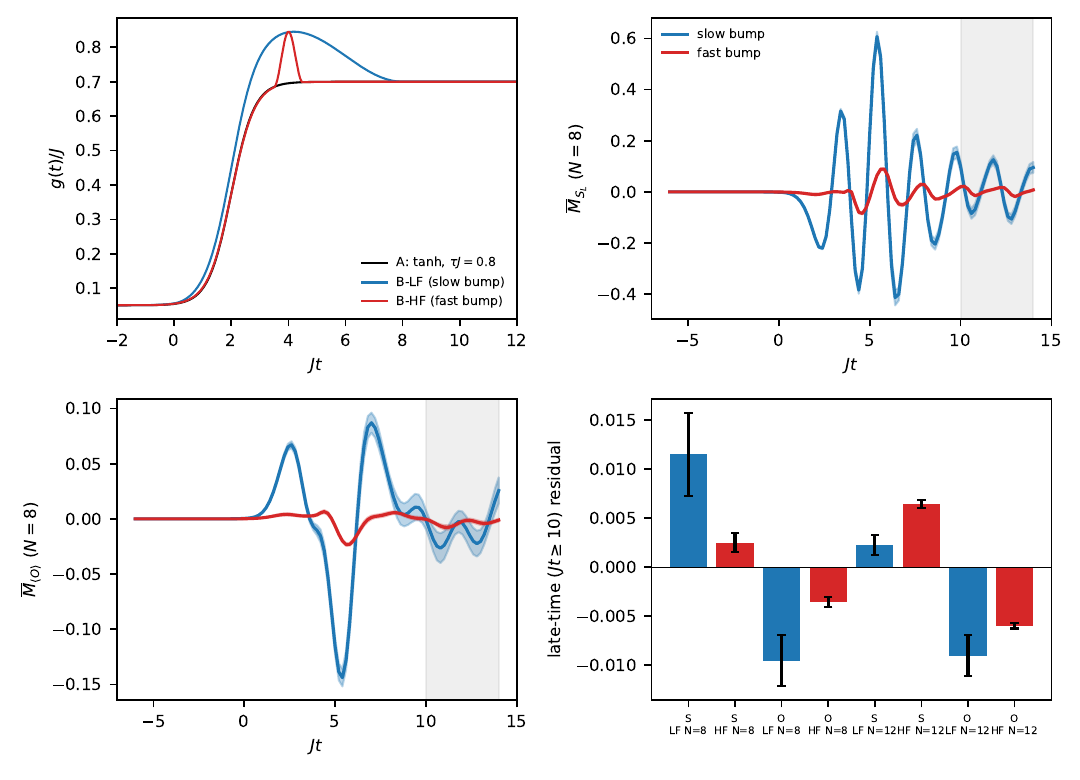}
\caption{Equal-work protocols that differ at low or at high frequency. Top left: $g(t)$ for the reference A and for B-LF and B-HF (at their mean tuned $\tau_B$). Top right and bottom left: memory functionals of the entropy and of $\langle\OO\rangle$ at $N=8$ ($30$ realizations; bands $\pm1$ SEM); the shaded region is the averaging window. Bottom right: late-time residuals at $N=8$ and $12$.}
\label{fig:tuned}
\end{figure}

\emph{Protocols.} The reference protocol A is the tanh switch of Eq.~\eqref{eq:protocol} with $\tau J=0.8$. The comparison protocol B is a tanh switch of duration $\tau_B$ plus a smooth bump $b(t)=a\sin^2[\pi(t-t_s)/L]$ on $t_s<t<t_s+L$, with $a=0.15J$. It has the same initial and final couplings as A, since the bump vanishes outside its support. In protocol B-LF the bump is slow, $L=8J^{-1}$ starting at $Jt_s=0$; in protocol B-HF it is fast, $L=1J^{-1}$ starting at $Jt_s=3.5$; both are centered at $Jt=4$ (Fig.~\ref{fig:tuned}, top left). The difference $g_B-g_A$ is therefore concentrated at frequencies $\lesssim2\pi/L\approx0.8J$ for B-LF and $\lesssim6J$ for B-HF. For each disorder realization the duration $\tau_B$ is tuned so that B deposits exactly the same work as A (matching on the exact energy change; maximum residual $4.2\times10^{-5}$ at $N=8$ and $5.4\times10^{-5}$ at $N=12$). The mean tuned durations are $JT_B=0.85$ (LF) and $0.82$ (HF) at $N=8$, and $0.88$ and $0.82$ at $N=12$.

\emph{Results.} Table~\ref{tab:tuned} gives the late-time residuals, averaged over $10\le Jt\le14$, and Fig.~\ref{fig:tuned} shows the time dependence. The residuals are small, between about $2\%$ and $11\%$ of the unmatched entropy difference of Table~\ref{tab:mem}, but they are not consistent with zero. They are \emph{not} systematically larger for the slow bump: at $N=8$ the entropy residual is larger for B-LF, but at $N=12$ it is larger for B-HF ($0.0064\pm0.0004$ against $0.0023\pm0.0010$); for $\langle\OO\rangle$ the two are comparable within about $1.5$ standard errors at $N=12$. The time dependence in Fig.~\ref{fig:tuned} is oscillatory and the residual at $N=8$ changes sign when the averaging window starts at $Jt=8$ instead of $10$ ($-0.005\pm0.002$ against $+0.012\pm0.004$ for the LF entropy), so these residuals are transients still ringing at the end of the run rather than a stationary late-time memory.

\begin{table}[t]
\centering
\footnotesize
\setlength{\tabcolsep}{4pt}
\begin{tabular}{lcccc}
\toprule
 & \multicolumn{2}{c}{$N=8$ ($30$ samples)} & \multicolumn{2}{c}{$N=12$ ($12$ samples)}\\
 & $\Delta S_L$ & $\Delta\langle\OO\rangle$ & $\Delta S_L$ & $\Delta\langle\OO\rangle$\\
\midrule
B-LF (slow bump) & $0.0115\pm0.0043$ & $-0.0096\pm0.0026$ & $0.0023\pm0.0010$ & $-0.0090\pm0.0021$\\
B-HF (fast bump) & $0.0025\pm0.0010$ & $-0.0036\pm0.0005$ & $0.0064\pm0.0004$ & $-0.0060\pm0.0003$\\
\bottomrule
\end{tabular}
\caption{Late-time ($10\le Jt\le14$) memory functionals, A minus B, at equal work (mean $\pm$ SEM).}
\label{tab:tuned}
\end{table}

\emph{Interpretation.} If the residual were an infrared memory carried by the logarithmic susceptibility, we would expect it to be larger when the protocol difference lies at low frequency. We do not find this. The residuals are of similar size for slow and fast differences, they oscillate in time, and they are of the size one expects from the energy distribution rather than the mean energy being matched, since the two protocols are matched only in mean energy. We conclude that at these sizes there is no evidence of a frequency-specific memory. This is consistent with the absence of an infrared window in systems of $8$ to $12$ Majoranas per side, where even a bump of duration $8J^{-1}$ has frequency content comparable to the level spacing, and it leaves the large-$N$ question open.

%=====================================================================
\section{Discussion}
\label{sec:discussion}

\paragraph{What the analysis establishes.} The central analytic result is that, with independent disorder, the first-order response of two coupled SYK systems to a time-dependent link has no interaction dressing: the cross-correlation is the product of two exact equilibrium propagators and the switching profile, with a definite factor of $i$ that exact diagonalization confirms. The same structure fixes the susceptibility of the link operator, whose dimension of one half makes it relevant and gives it a logarithmic infrared term with a definite coefficient and phase. An exact finite-size test of the first-order response and numerical solutions of the large-$N$ Euclidean equations support these statements quantitatively, at the few-percent level that finite size and finite $\beta J$ allow. The same susceptibility fixes the structure of the work, and predicts a logarithmic dependence on the switching time in the conformal regime.

\paragraph{Free versus interacting shutters.} In a free Majorana shutter the Heisenberg equations close on a one-particle space, $\bm\chi(t)=O(t)\bm\chi(0)$ with $O(t)\in SO(2N)$, and a positive-frequency decomposition leads naturally to a Bogoliubov kernel. In the coupled SYK problem the equation of motion $\dot\chi^L_i=i[H_L,\chi^L_i]+g\chi^R_i$ has a commutator that generates cubic Majorana operators, and repeated commutation generates operator strings of growing length. The exact unitary therefore cannot be summarized by a single one-particle matrix. The natural hierarchy is quadratic shutter, interacting Keldysh response, and many-body operator growth; an effective anomalous kernel is a possible projection between the second and first levels and does not replace the exact dynamics. The finite-size entropies, which exceed the quadratic ceiling by factors of three to five, show how far the exact evolution departs from the quadratic picture.

\paragraph{What the equal-work results do and do not say.} At equal energy deposition the late-time entropy, the link expectation value and (at $N=8$) the cross-system OTOC are nearly independent of the opening protocol. This is expected if the late-time state is determined mostly by the energy, as in a thermalizing system, and it is what one finds in a small system whose scrambling time is short compared with the observation window. The small residuals that remain are not frequency specific (Sec.~\ref{sec:tuned}), oscillate in time, and are of the size expected from matching only the mean energy. They do not contradict an infrared memory at large $N$, where the logarithm of Eq.~\eqref{eq:chilog} produces a slowly decaying response, but our systems are too small to have the window of scales in which this could be seen. The work itself offers a cleaner large-$N$ handle: Eq.~\eqref{eq:Wlog} predicts that, in the conformal window, the work depends logarithmically on the switching time with a definite coefficient. A decisive test would compare protocols with identical initial and final couplings and identical work in a system where the slow modes are resolved, by a driven Keldysh solution at large $N$ or by larger exact or tensor-network simulations, and would test the logarithmic work law of Eq.~\eqref{eq:Wlog} directly.

\paragraph{Relation to wormholes and to nearly-AdS$_2$ descriptions.} Two coupled SYK systems with a bilinear coupling are the basis of the eternal traversable wormhole~\cite{MaldacenaQi}. In that construction the two Hamiltonians are copies of each other, so cross rungs exist and the low-energy dynamics are controlled by a shared Schwarzian mode~\cite{MSY}; the gap then follows a power of the coupling set by the Schwarzian rather than by simple dimensional analysis. Here the disorder is independent, the cross rungs vanish, and the first-order physics is that of a relevant perturbation of two independent conformal systems. We find a gap whose local scaling exponent (about $1.5$--$1.6$) is neither the dimensional estimate nor linear, and we make no claim about a gravitational dual. A driven boundary description of the slow-opening regime is a natural direction, but it would require solving the coupled Schwarzian and Kadanoff--Baym dynamics with the same contour prescription as here.

\paragraph{Limits of the present work.} The large-$N$ results are first order in the link and second order in the work. The second-order cross response, where the ladder enters, and the effective quasiparticle projection are formulated but not solved, and the logarithmic work law is a prediction that we could not test. The gap exponent is resolved only in a window of $g/J$ of about a factor of two, and the finite-size statistics at $N=12$ ($10$ to $13$ realizations) are modest.

%=====================================================================
\section{Conclusions}
\label{sec:conclusions}

We set out to ask whether a strongly interacting many-body system remembers the way a link to a second such system was opened. The picture that emerges is layered. At the level of exact finite-size dynamics, opening the link transfers parity and builds entanglement well beyond what any quadratic description allows, and the entanglement grows with system size at a rate close to the maximum permitted. Yet at equal injected energy the late-time state, and the cross-system scrambling diagnostic at the size where we could test it, hardly depend on the opening protocol, so in these small systems the memory is carried almost entirely by the energy.

The large-$N$ analysis explains where a genuine memory could come from, and why it is subtle. Because the two systems are statistically independent, the leading response of the cross-correlation is not dressed by interactions at all; it is a product of two equilibrium propagators and the switching function. The link itself has dimension one half and is therefore relevant, so its susceptibility acquires a logarithm in the infrared. Responses that grow logarithmically at low frequency are the natural carriers of memory, because a finite-time protocol only partly filters the soft modes. At the same time, the coupled equilibrium problem shows that the infrared is cut off by a gap that grows more slowly than the simplest dimensional estimate predicts, which limits the window in which the logarithm applies and sharpens the question of how it is reached.

Two further results sharpen the picture. The work done by the switch splits into a first-order part, set by the polarization of the state just before the opening and independent of how fast the opening is, and a second-order part that grows in magnitude as the opening slows; the conformal susceptibility predicts that, where the logarithm is operative, this dependence is itself logarithmic. And protocols that differ only at low frequency, once matched in work, leave residuals no larger than protocols that differ at high frequency, so the small systems give no frequency-specific signal. Both statements are consistent with the picture that infrared memory is a large-$N$ phenomenon that finite exact simulations are too small to resolve.

Three problems remain open, and the framework of this paper defines them. The second-order response, which is where the interacting ladder enters, has to be solved in the driven Keldysh theory. The conjectured effective quasiparticle kernel, whose infrared norm we have characterized, has to be derived or refuted from that solution. And the equal-work comparison has to be repeated in a regime with a genuine infrared window. Answering these would decide whether the logarithm of the junction susceptibility translates into a protocol-dependent memory that survives equal work, which is the question this paper leaves precisely posed.

%=====================================================================
\appendix

\section{Keldysh structure}
\label{app:keldysh}

After the classical--quantum rotation, the contour Green function becomes the $2\times2$ matrix
\begin{equation}
\widehat G=\begin{pmatrix}G^K&G^R\\ G^A&0\end{pmatrix},
\end{equation}
and likewise for the self-energy, so that the Dyson equation reads $\widehat G^{-1}=\widehat G_0^{-1}-\widehat\Sigma_{\rm SYK}-\widehat\Sigma_{\rm link}$. The $LR$ block is generated by the matrix product $\delta\widehat G_{LR}=\widehat G\circ\widehat\Sigma_{{\rm link},LR}\circ\widehat G+O(g^2)$, and the $R$, $A$ and $K$ components follow from Keldysh matrix multiplication. A single scalar function therefore cannot be identified with the full nonequilibrium Green function; Proposition~\ref{prop:leading} is stated for the corresponding component.

\section{Numerical methods}
\label{app:numerics}

\paragraph{Exact dynamics.} The $\gamma_i$ are Jordan--Wigner matrices on $N$ qubits, with the first $N/2$ qubits belonging to the left system and the others to the right. The quartic terms of $H_L$ and $H_R$ are even and act only on their own qubits; $\OO$ contains the Jordan--Wigner strings. The couplings are Gaussian with variance $3!J^2/N^3$. States are evolved with a fourth-order Runge--Kutta scheme applied to $H(t)=H_0+g(t)\OO$ with a sparse $H_0$. The propagator $U(t)$ is evolved in the two blocks of total parity, which decouple because $\OO$ preserves it. The entropy is computed from the singular values of the state reshaped into a $2^{N/2}\times2^{N/2}$ matrix. The initial state uses the lowest even-parity eigenstate of each system, obtained by exact diagonalization of the single-system Hamiltonian in the even-parity block.

\paragraph{Exact test of Proposition~\ref{prop:leading}.} The retarded cross-correlation
\begin{equation*}
G^R_{LR,ik}(t,t')=-i\theta(t-t')\,\langle\{\chi^L_i(t),\chi^R_k(t')\}\rangle
\end{equation*}
is computed from the stored propagators $U(t_k)$ as $2\,\mathrm{Re}\langle w_k|v_l\rangle$ with $w_k=\chi_i^L(t_k)|\Omega\rangle$ and $v_l=\chi_k^R(t_l)|\Omega\rangle$ in the Heisenberg picture. The prediction uses the exact retarded matrices of the uncoupled systems and trapezoid quadrature in the intermediate time.

\paragraph{Euclidean large-$N$ solver.} The equations $G(i\omega)=1/(-i\omega-\Sigma(i\omega))$ and $\Sigma(\tau)=J^2G(\tau)^3$, or Eq.~\eqref{eq:GLLeq} at $g\neq0$, are solved on a midpoint $\tau$ grid with $2^{16}$--$2^{17}$ points by fast Fourier transform. The free tail $1/(-i\omega)$ is subtracted analytically, so that $G(\tau)=\tfrac12+\beta^{-1}\sum_n e^{-i\omega_n\tau}[G(i\omega_n)-1/(-i\omega_n)]$. Iteration uses damped mixing with an adaptively reduced weight until the maximum change is below $10^{-9}$--$10^{-11}$. The gap is the decay rate of $G_{LL}(\tau)$ between $\tau=\beta/8$ and $\beta/4$.

\paragraph{Kramers--Kronig and quadrature checks.} The sign of the susceptibility is verified by evaluating the Kramers--Kronig integral of a constant spectral weight on $|\omega|<J$. The Hilbert--Schmidt integrals are evaluated by adaptive two-dimensional quadrature.

\section{The Hilbert--Schmidt coefficient}
\label{app:hs}

Substituting $\omega=\mu x$, $\nu=\mu y$ and then $x=u^2$, $y=v^2$ in Eq.~\eqref{eq:HS} gives
\begin{equation}
C=\int_1^\infty\!\!\dd x\int_1^\infty\!\!\dd y\,\frac{1}{\sqrt{xy}\,(x+y)^2}=4\int_1^\infty\!\!\dd u\int_1^\infty\!\!\dd v\,\frac{1}{(u^2+v^2)^2}.
\end{equation}
In polar coordinates $u=r\cos\theta$, $v=r\sin\theta$ the region $u,v\ge1$ is $r\ge\max(\sec\theta,\csc\theta)$, and by reflection symmetry
\begin{equation}
C=8\int_0^{\pi/4}\dd\theta\int_{\csc\theta}^\infty\frac{\dd r}{r^3}=4\int_0^{\pi/4}\sin^2\theta\,\dd\theta=\frac\pi2-1 .
\end{equation}
The finite upper cutoff $\omega_\ast$ changes the integral by a term of order $\omega_\ast^{-1}$, from the region where either variable exceeds $\omega_\ast/\mu$.

\end{document}